\documentclass[11pt,letterpaper]{JHEP3}
\usepackage{graphics}
\usepackage{epsfig}
\usepackage{subfigure}
\usepackage{mathrsfs}
\usepackage{amsmath}
\usepackage{amsfonts}
\usepackage{amssymb}
\usepackage{graphicx}
\usepackage{longtable}
\usepackage{rotating}
\usepackage{multirow}
\title{Holographic thermalization and gravitational collapse in the spacetime dominated by quintessence dark energy}
\author{Xiao-Xiong Zeng\\
School of Science, Chongqing Jiaotong University, Chongqing, 400074, China\\
 \email{xxzeng@mail.bnu.edu.cn}}

\author{De-You Chen\\
Institute of Theoretical Physics, China West Normal University, Nanchong 637009, China\\
\email{deyouchen@126.com}
}

\author{Li-Fang Li\\
State Key Laboratory of Space Weather,
Center for Space Science and Applied Research, Chinese Academy of Sciences,
Beijing, 100190, China\\
\email{lilf@itp.ac.cn}
}
 \abstract{In this paper, the thermalization has been studied holographically. Explicitly in the gravity side, we consider the gravitational collapse of a thin shell of dust in a spacetime dominated by quintessence dark energy. With the thermalization probes such as the normalized geodesic length and minimal area surface, we study the effect of the state parameter for the quintessence dark energy on the thermalization. Our results show that the smaller the state parameter of quintessence is, the harder the plasma to thermalize. We also investigate the thermalization velocity and thermalization acceleration. We hope our results here can shed light on the nature of the quintessence dark energy.
}

\preprint{}
\begin{document}
\bibliographystyle{ieeetr}
\bibliography{reference.bib}

\section{Introduction}

Dark energy has been the hot topic since the discovery of the acceleration expansion of the universe. The famous scalar field models of dark energy include quintessence \cite{Wetterich:1987fm, Ratra:1987rm}, K-essence
\cite{ArmendarizPicon:2000dh, ArmendarizPicon:2000ah}, tachyon \cite{Padmanabhan:2002cp, Sen:2002in}, phantom \cite{Caldwell:1999ew, Nojiri:2003vn, Nojiri:2003jn} and so forth. For a comprehensive revision on dark energy models, see~\cite{Copeland:2006wr, Li:2011sd}. The equation of state parameter, which is defined by the ratio of pressure and energy density, is one of the crucial parameters for characterizing the nature of the above dark energy models. Therefore, extraction of the information on equation of state parameter plays an important role in understanding dark energy. Unfortunately, we have not known what the dark energy actually is.

Recently, many new data sets are released. These data contain the first 15.5 months observational data from the Planck mission on the temperature power spectrum of CMB \cite{planck_fit}, and the new SNIa data from the Pan-STARRS observational project \cite{Panst1, Panst2} and so on. The high quality data give new limits on the equation of state parameter. For example, the Planck data favor more negative state parameter of dark energy. Apart from these observations, there are other ways to help us study the nature of dark energy. Though studying the effect of the equation of state parameter on the physical system may enhance our understanding of dark energy. For example, we can probe the effect of dark energy on the black hole physics, which may provide a new hint for us to research dark energy.

Some attempts have been done to construct black hole solutions with dark energy by using of dynamical scalar fields~\cite{pbh,Kiselev}. Actually, in the present of a dynamical scalar field, the field equations of Einstein gravity are complicated and it is difficult to obtain the black hole solution with the state parameter of dark energy. Fortunately, in~\cite{Chen:2008ra}, the authors present a metric solution describing a d-dimensional planar quintessence AdS black hole. With such solution, the quasinormal modes and Hawking radiation of the black holes surrounded by quintessence have been studied~\cite{Sb1,Sb2}, which can help us further understand the relationship between dark energy and black hole. Especially recently there are also some works to study the quintessence AdS black hole in the framework of holography \cite{Chen:2012mva}. At present, although the dual field theoretical interpretation about the quintessence is still unclear, with the further investigations, one can get better understanding. With this motivation, in this paper we will study holographic thermalization in the spacetime dominated by quintessence dark energy.

 Thermalization is a process that the plasma of quarks and gluons are formed  in heavy ion colliders such as the RHIC
and LHC, and it was shown that this process is strongly coupled.  Previously, one  treated usually this process as a near-equilibrium dynamics and  mapped it
onto the dynamics of perturbations of the $AdS_5$-black brane metric. Recent experiment datas in  RHIC  show that the time scale for equilibration of matter is  shorter than expected in the framework of perturbative approaches to thermalization, which motivates the use of the AdS/CFT correspondence to study thermalization of strongly coupled plasmas. Until now,  many  models have been constructed to study this  strongly coupled non-equilibrium  process~\cite{Garfinkle84, Garfnkle1202, Allais1201, Das343, Steineder, Wu1210, Gao, Buchel2013, Keranen2012, Craps1, Craps2,  Balasubramanian1, Balasubramanian2}.
Among them, one elegant model is presented in~\cite{Balasubramanian1, Balasubramanian2}, where an AdS-Vaidya metric that interpolates between pure AdS space and Schwarzschild-AdS black hole  was used to study the thermalization
of the boundary conformal field theory.
With the AdS/CFT correspondence in this case,  the AdS  space is dual to the initial state, the collapse of a thin shell of dust is dual to the sudden injection of energy, and the Schwarzschild-AdS black hole is dual to the final equilibrium state in the conformal field theory. To probe the thermalization process, the non-local observables such as two-point correlation function, Wilson loop, and entanglement entropy were used. In the saddle approximation, these probes are dual to the geodesic, minimal surface, and minimal volume in this bulk. Based on this method, the holographic thermalization investigation has been generalized to the bulk geometry with electrostatic potential~\cite{GS,CK, Yang1}, high curvature corrections~\cite{Zeng2013, Zeng2014, Baron, Li3764}, and other gravity models~\cite{Baron1212,Arefeva,Hubeny,Arefeva6041,Balasubramanianeyal6066,Balasubramanian4,Balasubramanian3,Balasubramanian9,Fonda,Cardoso,Hubeny2014,Zeng2015}.

The purpose of this  paper is to investigate the effect of the state parameters for quintessence dark energy on the holographic thermalization. We take the
two-point correlation function and expectation value of Wilson loop as thermalization probes. For each thermalization probes, we study the evolution of the renormalized observables. We find for both the thermalization probes, smaller the
state parameter of quintessence is, longer the thermalization time is\footnote{The state parameters of the quintessence vary in the range  $-1\leq w<0$, the large state parameters are that near $0$ and the small   state parameters are that near $-1$. }. That is, as the state parameter decreases, the plasma in the dual conformal field theory is harder to thermalize. In addition, we find the state parameter has an important effect on the thermalization probes. Explicitly, for the large state parameters, we find there is a slight delay in the onset of thermalization for both the thermalization probes, which is similar to that in the Vaidya AdS black branes~\cite{Balasubramanian1, Balasubramanian2}. After this delay, there is a quadratic growth and linear growth stage as found in~\cite{liu1, liu2}.
While for a small state parameter, we find the delay becomes shorter and to some case, the renormalized observables grow immediately. To understand the thermalization profoundly, we also obtained the thermalization velocity and thermalization acceleration. As the state parameter of quintessence decreases, the initial values for both the thermalization probes increase from negative to positive. Especially for the large state parameters, there is a phase transition point during the whole thermalization process for both thermalization probes, which divides the thermalization into an acceleration phase and a deceleration phase. We further investigate the effect of the state parameters on the phase transition point and find that the phase transition point decreases as the state parameters decrease. But for a small state parameters, the thermalization velocity is found to be decreased during the whole thermalization process for the initial thermalization velocity which is too large. The thermalization acceleration, of course, is negative always in this case.

This paper is organized as follows. In the next section, we will briefly review the gravitational collapse solution in the spacetime dominated by the quintessence dark energy. Then in Section~\ref{Nonlocal_observables}, we derive the equations of motion of the non-local observables theoretically. In Section~\ref{Numerical_results}, we perform a systematic analysis about how the state parameters affect the thermalization time resorting to numerical calculation. We also study the thermalization velocity and thermalization acceleration by the fitting functions of the thermalization probes. The last section is devoted to discussions and conclusions.

\section{The quintessence Vaidya AdS black branes}
\label{quintessence_Vaidya_AdS}
In this section, we will first review the AdS black brane solution dominated by quintessence and then extend it to the gravitational collapse case. The quintessence AdS black brane is also the solution of the Einstein field equation
\begin{equation}
R_{\mu\nu}-\frac{1}{2}Rg_{\mu\nu}-\frac{(D-1)(D-2)}{2L^2}  g_{\mu\nu} = 8 \pi G T_{\mu\nu}^{q},  \label{equation}
\end{equation}
where $L$ is the radius of the AdS, $D$ is the dimension of the spacetime, and $ T_{\mu\nu}^{q}$ is the energy-momentum tensor
of the quintessence whose nonzero components are given as
\begin{eqnarray}
 &T_{t}^{qt}&= T_{r}^{qr}=-\rho_q,  \nonumber \\
 &T_{x_1}^{qx_1}&= T_{x_2}^{qx_2}=\cdots =T_{x_{(D-2)}}^{qx_{(D-2)}}= \frac{\rho_q}{D-2} [(D-1)w+1],   \label{tensor}
\end{eqnarray}
in which $\rho_q $ is the dark energy density, and $w$ is the state parameter of quintessence.
Combing Eq.(\ref{equation}) with Eq.(\ref{tensor}), one can  get a
static spherically symmetric quintessence AdS black brane~\cite{Chen:2012mva}
\begin{equation}
 ds^{2}=-f(r)dt^{2}+f^{-1}(r)dr^{2}+r^{2}dx_i^2,   i=1,2,\cdots D-2 ,\label{areacoorections1}
\end{equation}
where%
\begin{equation}
 f(r)=\frac{r^2}{L^2}+\frac{c_1}{r^{D-3}}+\frac{c_2}{r^{(D-1)w+D-3
}}, \label{areacoorections2}
\end{equation}%
in which $c_1$ and $c_2$ are normalization factors. The energy density $\rho_q $ of quintessence and the state parameter $w$
have the following relation
\begin{equation}
\rho_q =\frac{c_2(D-1)(D-2)w}{2 r^{(D-1){(w+1)}}}.
\end{equation}
The background~Eq.(\ref{areacoorections2}) admits many spacetimes with different choices of $c_1$ and $c_2$. In general, for the usual quintessence, the energy density $\rho_q $ is positive and the state parameter $w$ is negative,
which means that the constant $c_2$  must be negative. Here we will limit us to the case $c_1=0$ and $c_2=-M$ as done in~\cite{Chen:2012mva} for simplicity, where $M$ is called as the ¡°quintessence charge¡±. With this assumption, Eq.(\ref{areacoorections2}) reduces to
\begin{equation}
 f(r)=\frac{r^2}{L^2}-\frac{M}{r^{(D-1)w+D-3
}}.  \label{areacoorections22}
\end{equation}%
The temperature of the quintessence AdS black brane in this case can be calculated to be
\begin{equation}
T_{q}=\frac{(D-1)(w+1)r_h}{4\pi L^2}, \label{temperature}
\end{equation}
where $r_h=(ML^2)^{1/(D-1)(w+1)}$ is the event horizon of the black brane and determined by $f(r_h)=0$.
As $w \rightarrow 0$, the background we are interested
reduces to the usual planar Schwarzschild AdS black brane, while $w \rightarrow -1$ it reduces to a pure AdS space with a renormalized cosmological
constant $1/L^2-M$.

To get the quintessence Vaidya black brane solution, we make the coordinate
transformation $z=\frac{L^2}{r}$, with which the black brane metric in Eq.(\ref{areacoorections1}) can be cast into
\begin{equation}
 ds^{2}=\frac{1}{z^2}[-H(z)dt^{2}+H^{-1}(z)dz^{2}+dx_i^2], \label{metric}
\end{equation}
\begin{equation}
 H(z)=1-z^{(D-1)w+D-1}M, \label{h}
\end{equation}
where $L$ has been set to $1$.

Introducing the Eddington-Finkelstein coordinate
\begin{equation}
dv=dt-\frac{1}{H(z)}dz,
\end{equation}
the background spacetime in  Eq.(\ref{metric}) then changes into
\begin{equation}
ds^2=\frac{1}{z^2} \left[ - H(z) d{v}^2 - 2 dz\ dv +
dx_i^2 \right]. \label{collpse}
\end{equation}
The quintessence Vaidya AdS black brane can be obtained by
freeing the quintessence charge in Eq.(\ref{h}) as an arbitrary function of $v$. In this case, Eq.(\ref{collpse}) can be treated as the solution of the following field equation
\begin{equation}
R_{\mu\nu}-\frac{1}{2}Rg_{\mu\nu}-\frac{(D-1)(D-2)}{2L^2}  g_{\mu\nu} = 8 \pi G (T_{\mu\nu}^{q}+T_{\mu\nu}^{m}),
\end{equation}
where
\begin{equation}
T_{\mu\nu}^m\propto(D-2)z^{D-2}\frac{dM(v)}{dv}\delta_{\mu v}\delta_{\nu v},
\end{equation}
in which  $M(v)$ is the quintessence charge of a collapsing quintessence black brane.

Up to now, the gravitational collapse solution has been constructed. Next we are going to investigate the holographic thermalization process in the gravitational background~(\ref{collpse}). According to the AdS/CFT correspondence, the initial state in the conformal field theory is dual to the AdS boundary in a higher dimensional spacetime, the sudden injection of energy is dual to the collapse of a thin shell of dust, and the final equilibrium state is dual to a static black brane. In order to describe the thermalization process holographically, one should construct a proper model in the bulk. We set the quintessence charge as $M(v)=M\eta({v})$, where $\eta({v})$ is the step function. More explicitly, $M(v)$ is often written as the smooth function
\begin{equation}
M(v) = \frac{M}{2} \left( 1 + \tanh \frac{v}{v_0} \right),
\end{equation}
where $v_0$ represents the thickness of this finite shell. In the limit $v\rightarrow-\infty$, the background corresponds to a pure AdS space which is dual to the vacuum state in the conformal field theory, while in the limit $v\rightarrow \infty$, it corresponds to a quintessence AdS black brane which is dual to the equilibrium state in the conformal field side.

\section{Nonlocal observables}
\label{Nonlocal_observables}
In this section, we describe the thermalization holographically. As stated before, the probes in the gravity side are given by the area of the bulk extremal surface for a given configuration. Without loss of generality, we will consider an $n$-dimensional strip $\Sigma$ which ends at an $(n-1)$-dimensional spatial surface $\partial \Sigma $ lying at some time in the boundary theory. In this case, the strip consists of
two $(n-1)$- dimensional hyperplanes located at\cite{liu1, liu2}
\begin{equation}
\Sigma_n: -l/2\leq x_1\leq l/2; 0\leq x_2\leq\cdots \leq x_n \leq R_0; x_a\equiv 0, a=n+1, \cdots D-2,
\end{equation}
in which $0 < R_0<\infty$. For $n=1$, $\Sigma$ consists of two points separated by $-l/2\leq x_1\leq l/2$,
the extremal surface is the geodesic connecting
the two points, and its length $L$ gives
the equal-time two point correlation function of an
operator with large dimension\cite{Balasubramanian61}
\begin{equation}
\langle {\cal{O}} (t_0,x_i) {\cal{O}}(t_0, x_j)\rangle  \approx
e^{-\Delta {L}} ,\label{llll}
\end{equation}
where $\Delta$ is the conformal dimension of scalar operator $\cal{O}$.
For $n=2$, $\Sigma$ is a closed line, which defines the contour of a spacelike Wilson loop, and the area of the extremal surface  gives the expectation value of the Wilson loop operator~\cite{Maldacena80}
\begin{equation}
\langle W_{\Sigma}(t)\rangle \approx e^{-\frac{A_{\Sigma}(t)}{2\pi\alpha'}},
\end{equation}
where $\alpha'$ is the Regge slope parameter.
While for $n=D-2$, $\Sigma$ is a closed surface
which separates space into two regions. The area
$A_{\Sigma}$ then gives the entanglement entropy associated
with the region bounded by $\Sigma$ \cite{Ryu181602, Hubeny0707}
\begin{equation}
S_{\Sigma}(t)=\frac{A_{\Sigma}(t)}{4 \pi G_D},
\end{equation}
where $G_D$ is the $D$-dimensional Newton's constant. It is obvious that to get the equation of motions for non-local observables, one should get the area of the extremal surface ending
on the strip $\Sigma$
\begin{equation}
A_n=\int \sqrt{g_{\mu\nu}dx^{\mu}dx^{\nu}}dx\int dx_{\delta},
\end{equation}
 here for notational simplicity, $x_1$ has been
replaced by $x$ and  $\delta=2,3,\cdots,n$.
 For the quintessence Vaidya AdS black branes, the area of an $n$-dimensional surface  can be written as
\begin{eqnarray}
A_n&=& \frac{R_0^{n-1}}{z^{n-1}}\int_{-\frac{l}{2}}^{\frac{l}{2}} \frac{\sqrt{\Pi}}{z}\nonumber\\
   &=& K_n \int_0^{\frac{l}{2}} dx \frac{\sqrt{\Pi}}{z^n} ,\label{false}
\end{eqnarray}
with
\begin{equation}
\Pi=1-2z'(x)v'(x) - H(v,z) v'(x)^2,~~ K_n=2R_0^{n-1},
\end{equation}
Note that the integrand in Eq.(\ref{false}) can be thought of as the Lagrangian $\cal{L}$ of a fictitious system with $x$ the proper time. So according to the Lagrangian equation
\begin{equation}
\frac{\partial}{\partial x}(\frac{\partial \cal{L}}{\partial q'_i})=\frac{\partial \cal{L}}{\partial q_i}, i=1,2,
\end{equation}
where $q_1$, $q_2$ denote $z(x)$  and  $v(x)$ respectively. We can get the equations of motion for $z(x)$  and  $v(x)$
\begin{equation}\label{lequation}
z^n(x)\sqrt{\Pi}\partial_x (\frac{z'(x)+H(v,z)v'(x)}{z^n(x)\sqrt{\Pi}})=\frac{1}{2}\frac{\partial H(v,z) }{\partial v(x)} v'^2(x),
\end{equation}
\begin{equation}\label{aequation}
z^n(x)\sqrt{\Pi}\partial_x (\frac{v'(x)}{z^n(x)\sqrt{\Pi}})=\frac{1}{2}\frac{\partial H(v,z) }{\partial z(x)} v'^2(x)+n \frac{\Pi}{ z(x)}.
\end{equation}
Furthermore, by the reflection symmetry of the strip, we have the following initial conditions
\begin{equation}\label{initial}
z(0)=z_{\star},  v(0)=v_{\star} , v'(0) =
z'(0) = 0.
\end{equation}
With them, the configuration of bulk extremal surfaces will be fully fixed for a given $n$.
Since the Lagrangian does not depend explicitly on $x$, there is an associated conserved quantity
\begin{equation}
\sqrt{\Pi}z^n=z_{\star}^n,
\end{equation}
according to this relation, Eq.(\ref{false}) can be simplified as
\begin{equation}
A_n=K_n \int_0^{l/2} dx \frac{z_{\star}^n}{z^{2n}}.
\end{equation}
Generically the area of the extremal surface is divergent, so one needs to make regularization, which is achieved by
imposing the boundary conditions as follows
\begin{equation}\label{regularization}
z(\frac{l}{2})=z_0, v(\frac{l}{2})=t_0,
\end{equation}
where $z_0$ is the IR radial cut-off and $t_0$ is the time that the extremal surface approaches to the boundary, which is called thermalization time. In this paper, we are interested in the case $D=4$ for there have been many works to study the effect of
the spacetime dimensions on the thermalization probes \cite{Balasubramanian1, Balasubramanian2, GS, CK}. According to the definition of the area of the extremal surface, we know that the entanglement entropy equals to the expectation value of the
Wilson loop, thus next, we will use the two point correlation function and Wilson loop to explore effect of quintessence on the thermalization. In the bulk, this behavior equals to probe the gravitational collapse in a spacetime surrounded by quintessence using the geodesic length and minimal area surface, for which the renormalized forms are
\begin{equation}\label{lren}
\delta L=2 \int_{0}^{\frac{l}{2}} dx \frac{z_{\star}}{z(x)^2}-2\ln(2/ z_0),
\end{equation}
\begin{equation}\label{aren}
\delta  A=2\int_0^{\frac{l}{2}}dx
\frac{z^2_{\star}}{z(x)^4}-\frac{2}{z_0},
\end{equation}
where $R_0$ have also been set to $1$. Obviously as $z(x)$ in Eq.(\ref{lequation}) and  Eq.(\ref{aequation}) are solved for a given $n$, we can get the  renormalized  geodesic length and minimal area surface.

\section{Numerical results}
\label{Numerical_results}
In this section, we will first study how the quintessence affects the evolution of the geodesic and  minimal area surface by solving Eq.(\ref{lequation}) and  Eq.(\ref{aequation}) with the initial conditions in Eq.(\ref{initial}). We mainly concentrate on how the quintessence affects the thermalization time.
During the numerical simulation, we will take the shell thickness $v_0 = 0.01$ and UV cut-off $z_0=0.01$ respectively. The quintessence charge $M$ will be set to one so that the horizon is always located at $z_h=1$.

\subsection{Two point function}
\label{two_point}
\begin{figure}
\centering
\subfigure[$v_{\star}=-0.777, w=0$]{
\includegraphics[scale=0.5]{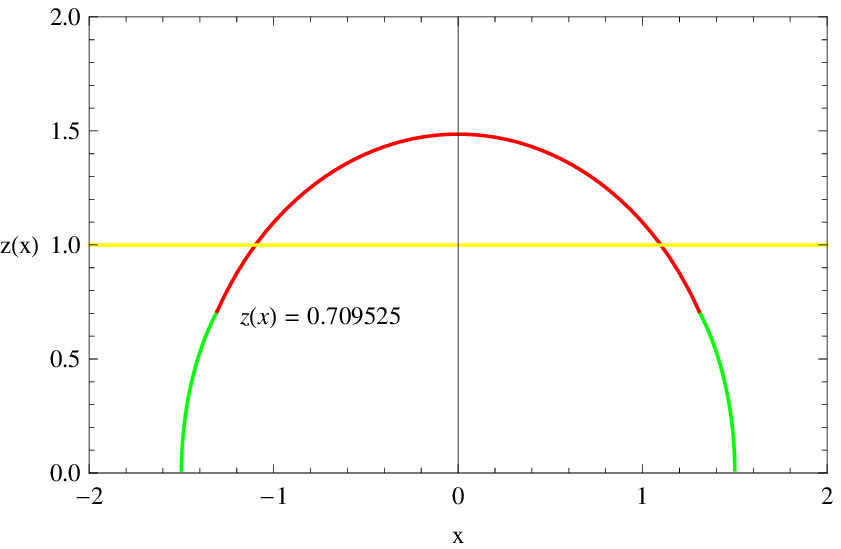} }
\subfigure[$v_{\star}=-0.444, w=0$]{
\includegraphics[scale=0.5]{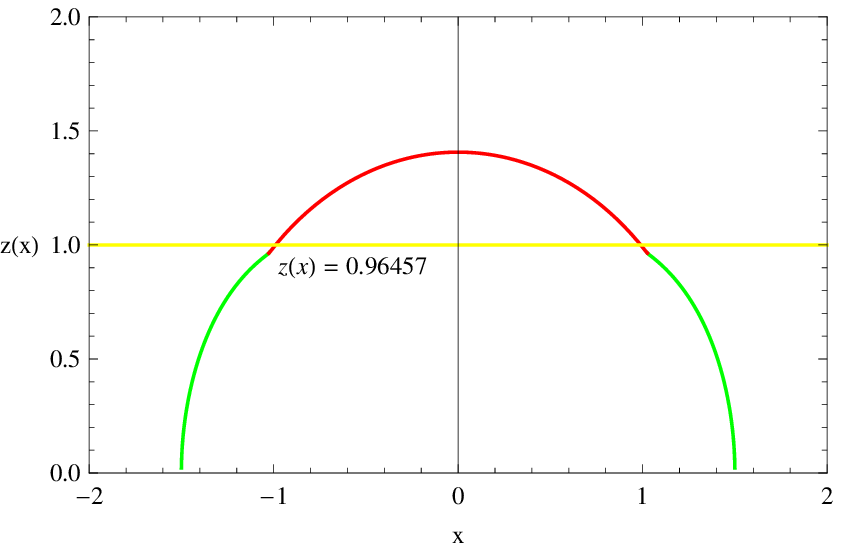}
}
\subfigure[$v_{\star}=-0.111, w=0$]{
\includegraphics[scale=0.5]{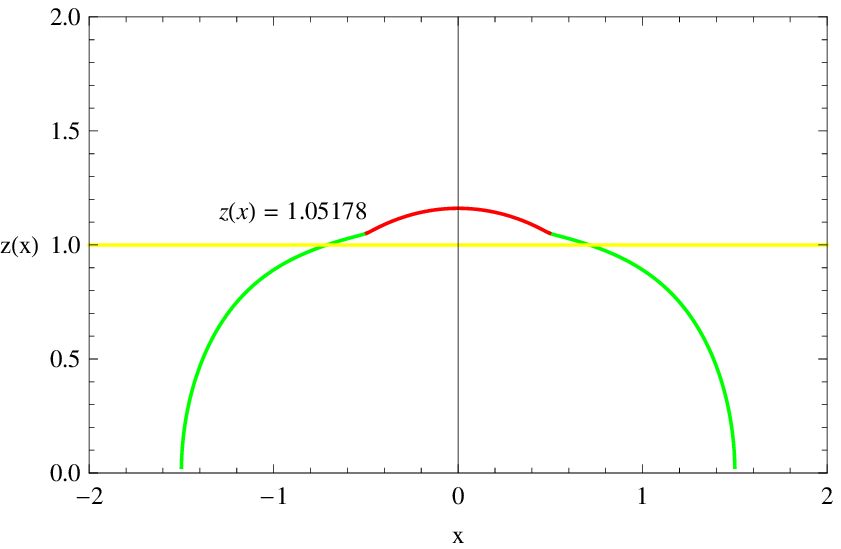}
 }
\subfigure[$v_{\star}=-0.777, w=-0.4$]{
\includegraphics[scale=0.5]{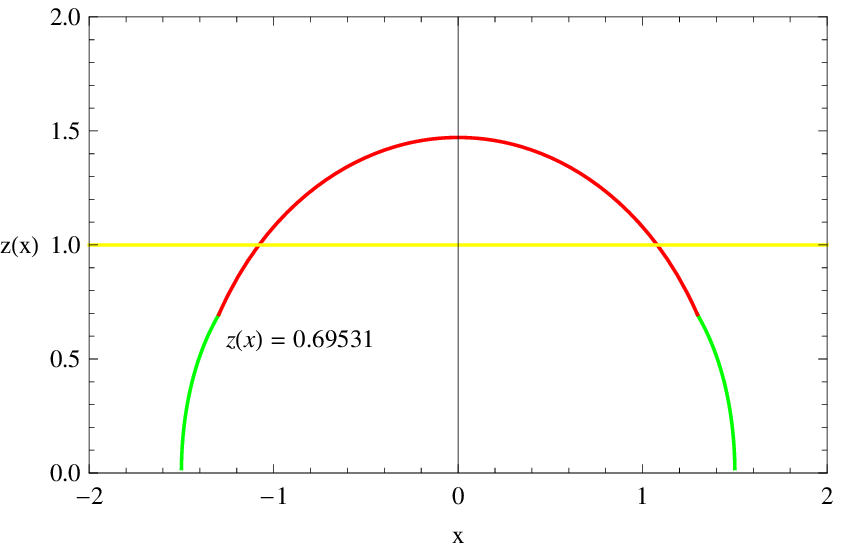}
}
\subfigure[$v_{\star}=-0.444, w=-0.4$]{
\includegraphics[scale=0.5]{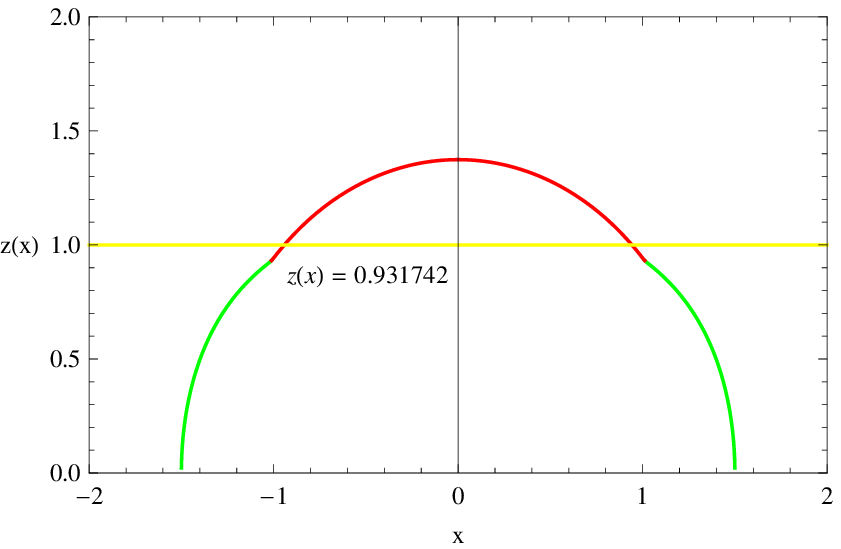}
}
\subfigure[$v_{\star}=-0.111,  w=-0.4$]{
\includegraphics[scale=0.5]{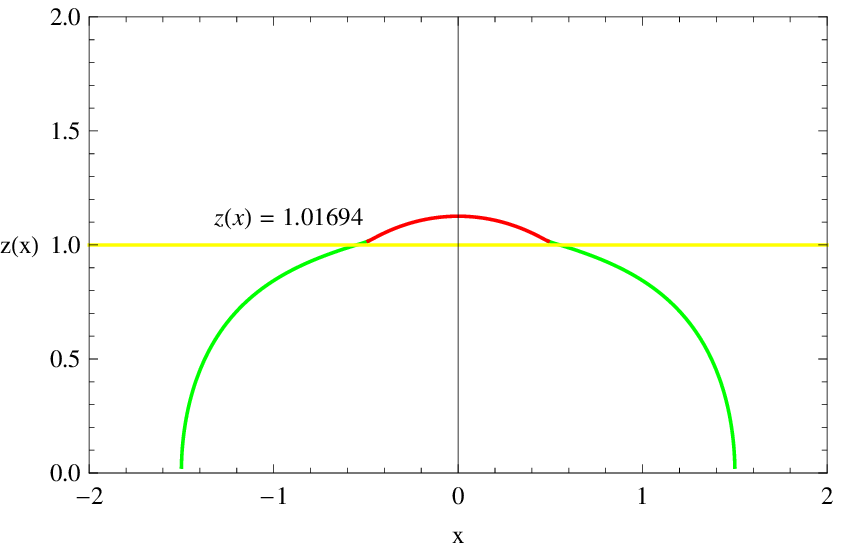}
 }
\subfigure[$v_{\star}-0.777, w=-0.8$]{
\includegraphics[scale=0.5]{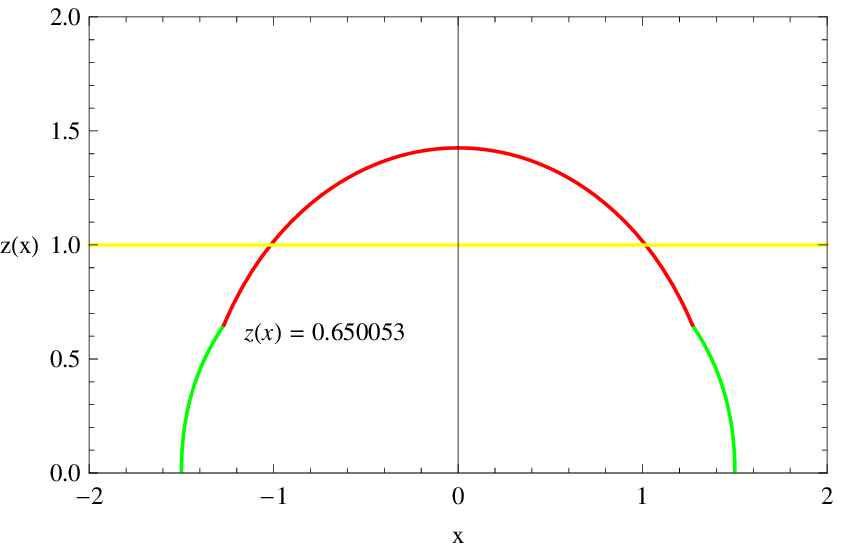}
}
\subfigure[$v_{\star}=-0.444, w=-0.8$]{
\includegraphics[scale=0.5]{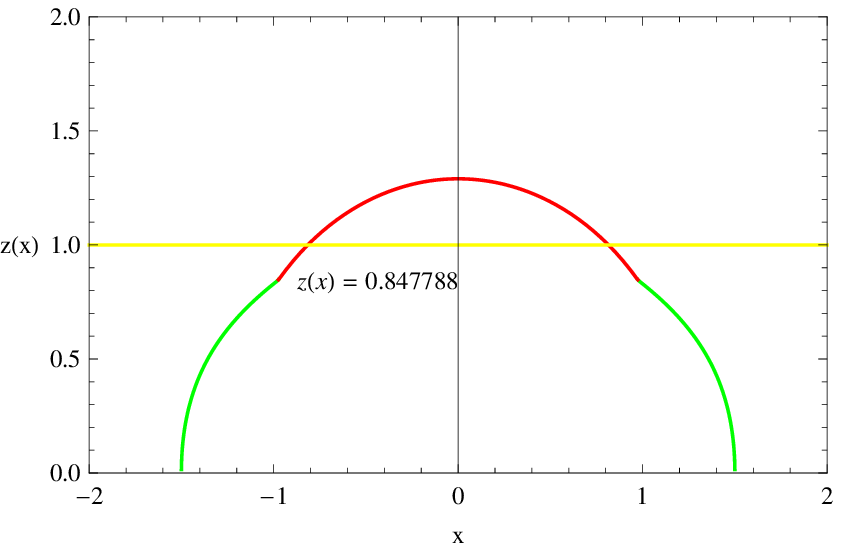}
 }
\subfigure[$v_{\star}=-0.111, w=-0.8$]{
\includegraphics[scale=0.5]{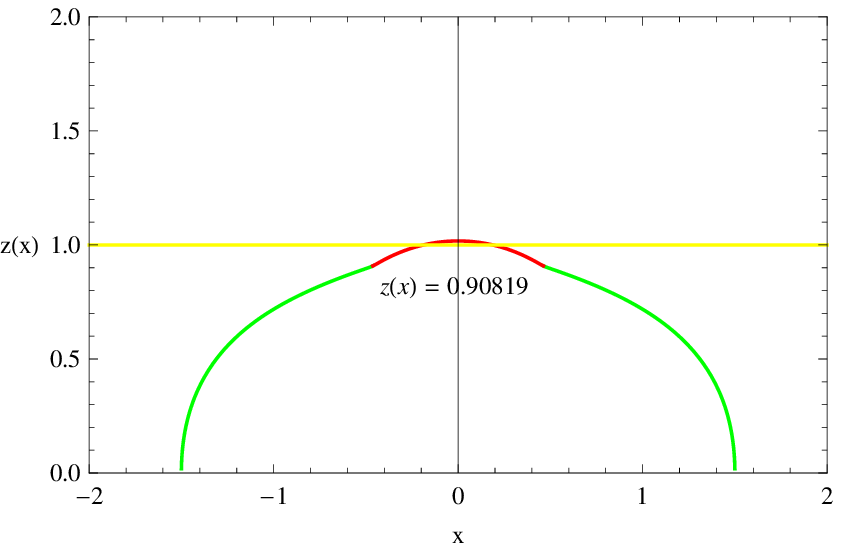}
 }
 \caption{\small Motion profile of the geodesics in the  quintessence Vaidya AdS black brane. The separation of the
boundary field theory  is $\l=3$. The black brane
horizon is indicated by the yellow line. The  position of the shell is described by the junction between the  red line and the green line.} \label{fig1}
\end{figure}

In this subsection, we probe the thermalization of the quark gluon plasma with two point function, which corresponds to $n=1$. In the bulk, it is dual to probe the gravitational collapse of a shell of dust by the geodesic. Our numerical results are shown in Table (\ref{tab1}). In Table (\ref{tab1}), we list the thermalization time for different state parameters at different initial time $v_{\star}$. It is easy to see that for a fixed initial time, as the state parameters decrease from zero to negative, the thermalization time increases step by step. Especially, as $w$ decreases from $-0.6$ to $-0.8$, the change of the thermalization time is most obvious. Therefore, we concluded that the smaller state parameters of quintessence delay the thermalization. This phenomenon also can be observed by studying the motion profiles of the geodesic, which are shown in Figure (\ref{fig1}). From Figure (\ref{fig1}), we know that for a fixed initial time $v_{\star}$, as the state parameter of quintessence $w$ decreases from zero to $-0.8$, the shell goes away from the horizon. This phenomenon is most obvious for the case $v_{\star}=-0.111$. In such case, we can see that for $w=0,  -0.4$, the shell has been dropped into the horizon while for the case  $w= -0.8$, it is out of the horizon. Thus a static black brane has been formed for $w=0, -0.4$ while the shell is collapsing for $w= -0.8$. In the dual conformal field theory, it means that the plasma for the case $w=0, -0.4$ have been thermalizaed while for $w= -0.8$, it is thermalizing, which also means that as the state parameter of quintessence decreases, the thermalization is delayed.

%======================figure1===================

\begin{table}
\begin{center}\begin{tabular}{l|c|c|c|c|c}
 %\MC{3}{c}{\text{caption}}\\[5pt]
 \hline
% & \multicolumn{3}{c||}{MGCDM}   & \multicolumn{3}{c}{$\Lambda$CDM}  \\ \hline
%                             &        MGCDM        &                  &             &      $\Lambda$CDM    &                   & \\ \hline
% \MC{3}{|c|c|}{\ZZ
%{15pt}\hfill\normalsize   \hfill  \hfill\normalsize MGCDM     \hfill\normalsize $\Lambda$CDM  }\\ \hline
% \ZZ{-6pt}{22pt}
               &$w=0$ &          $w=-0.2$  &       $w=-0.4$   &       $w=-0.6$  &   $w=-0.8$    \\ \hline
$v_{\star}$=-0.777    & 0.702861     &  0.712679      & 0.735064         & 0.788356         & 0.949775   \\ \hline
$v_{\star}$=-0.444    &0.995173    &1.02957          &1.0837         &1.18319      & 1.43399     \\ \hline
$v_{\star}$=-0.111     &1.27647      &1.33376       & 1.41476            &1.54978       &1.86324       \\ \hline
\end{tabular}
\end{center}
\caption{The thermalization time $t_0$ of the geodesic probe for different state parameters of quintessence $w$ and different initial time $v_{\star}$ with the same boundary separation $l=3$.}\label{tab1}
\end{table}

\begin{figure}
\centering
\subfigure[$l=2$]{
\includegraphics[scale=0.75]{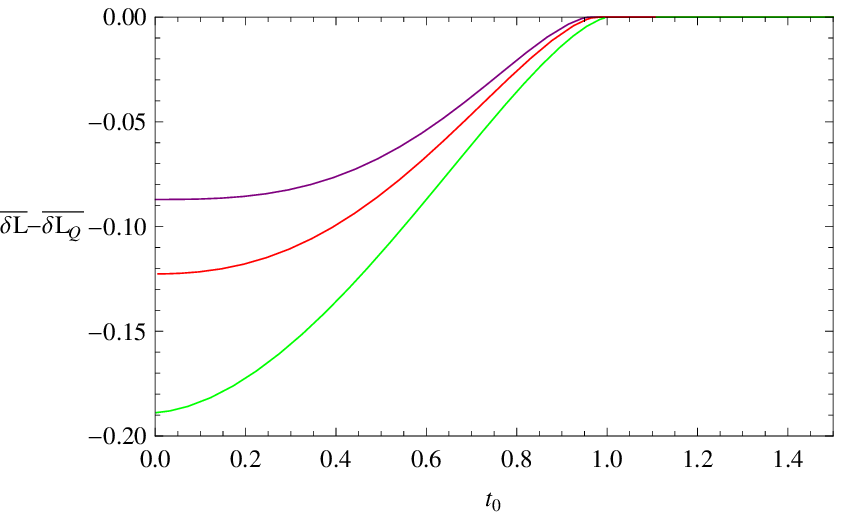}  }
\subfigure[$l=2$]{
\includegraphics[scale=0.75]{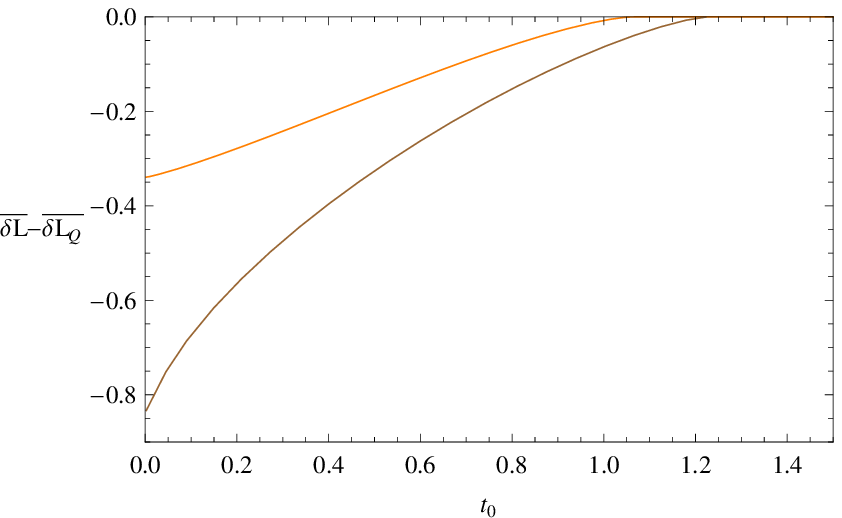}
}
 \caption{\small Thermalization of the renormalized geodesic length in a quintessence  Vaidya AdS  black brane for different state parameters at $l=2$.
 The purple line, red line,  and green line in  (a) correspond to  $w=0, -0.2, -0.4$  and the orange line, brown line  in  (b) correspond to  $w=-0.6, -0.8$  respectively.} \label{fig2}
\end{figure}

\begin{figure}
\centering
\subfigure[$l=3$]{
\includegraphics[scale=0.75]{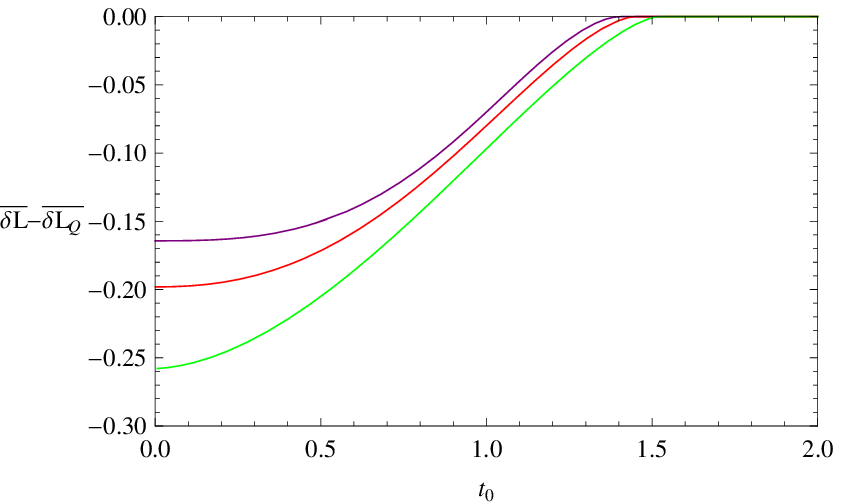}}
\subfigure[$l=3$]{
\includegraphics[scale=0.75]{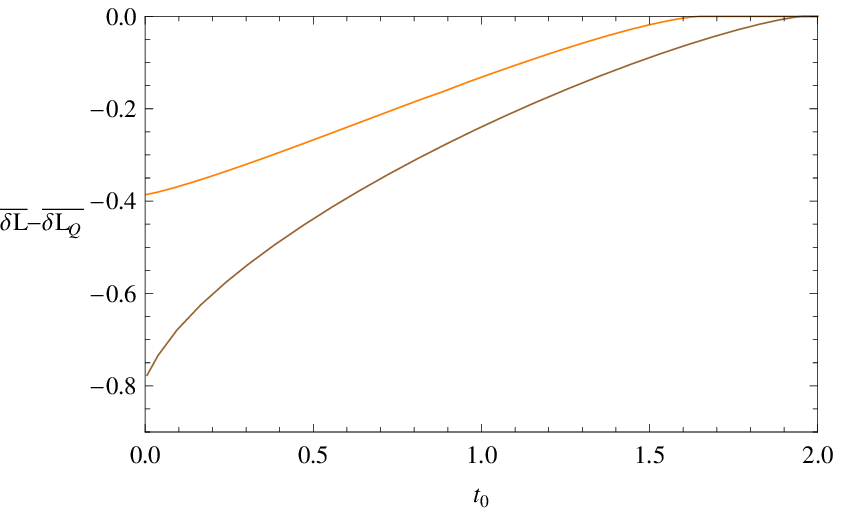}
}
 \caption{\small Thermalization of the renormalized geodesic length in a quintessence  Vaidya AdS  black brane for different state parameters at $l=3$.
 The purple line, red line,  and green line in  (a) correspond to  $w=0, -0.2, -0.4$  and the orange line, brown line  in  (b) correspond to  $w=-0.6, -0.8$  respectively.} \label{fig3}
\end{figure}

\begin{figure}
\centering
\subfigure[$l=3$]{
\includegraphics[scale=0.75]{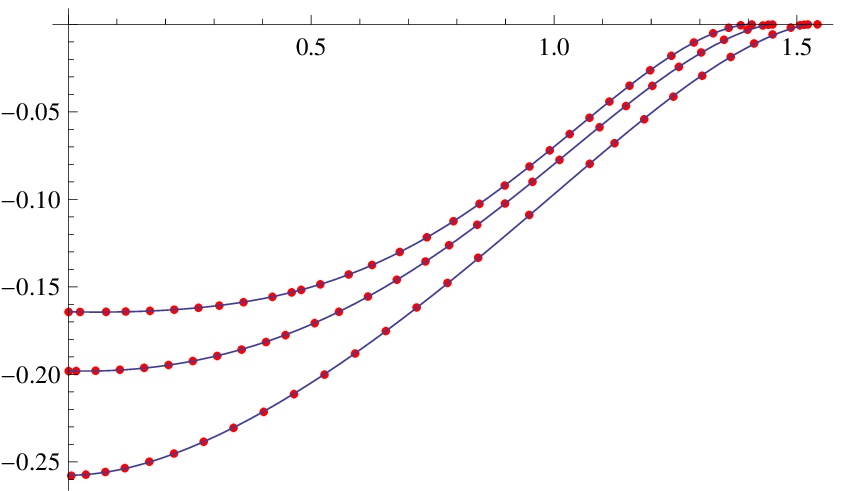} }
\subfigure[$l=3$]{
\includegraphics[scale=0.75]{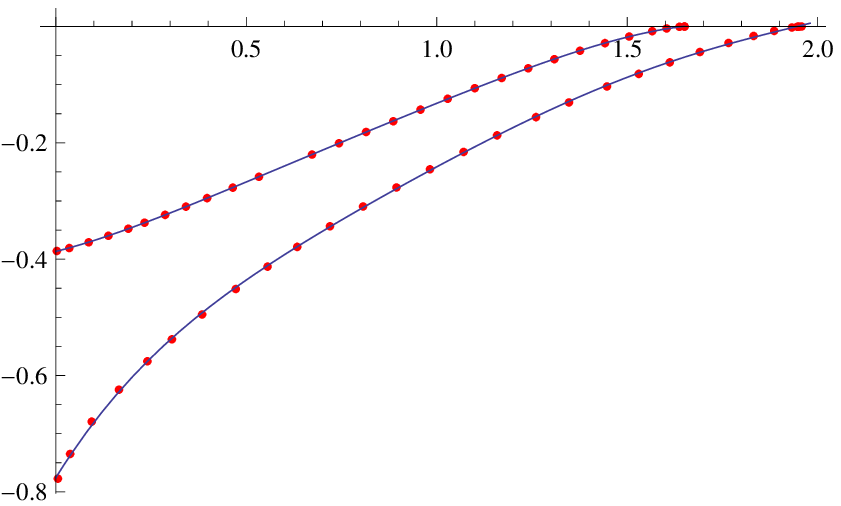}
}
 \caption{\small Comparison of the functions in Eq.(4.1) with the numerical results in Figure (3).} \label{fig4}
\end{figure}

We further study the renormalized geodesic length with the help of Eq.(\ref{lren}). As done in~\cite{Balasubramanian1, Balasubramanian2, GS}, we compare $\delta L$
at each time with the final value $\delta L_{Q}$, which is obtained in a static quintessence  AdS black brane, \emph{i.e.} $M(\mu)=M$. In this case, the thermalized state is labeled by the zero point of the vertical coordinate in each picture. We will plot the invariant quantity $\overline{\delta L}\equiv \delta L/l$ in Figure (\ref{fig2}) and Figure (\ref{fig3}), which give the relation between the  renormalized geodesic length and thermalization time for different state parameters $w$ at a fixed boundary separation. In each picture, the vertical axis indicates the renormalized geodesic length while the horizontal axis indicates the thermalization  time $t_0$. From Figure (\ref{fig2}) and Figure (\ref{fig3}), we know that at a fixed boundary separation, as the state parameter decreases, the thermalization time increases. This phenomenon is most obvious for small state parameters. Therefore, as the state parameter of quintessence decreases, the thermalization is delayed, which is consistent to our previous observation.

In addition, in Figure(\ref{fig2}) and Figure (\ref{fig3}), we find that for the large state parameters $w$, there is a delay in the onset of thermalization for the   renormalized geodesic length as shown in~\cite{Balasubramanian1, Balasubramanian2}. And after that stage, there is a quadratic growth and linear growth stage as found in \cite{liu1, liu2}. However especially for $w=-0.6, -0.8$, we find the evolution of the renormalized geodesic length has different tendency completely. For $w=-0.6$, the whole process of the evolution seems to be a linear growth, while for $w=-0.8$, it seems to be a high order growth. To understand profoundly how the quintessence dark energy affects the gravitational collapse process, next we will study the thermalization velocity and thermalization acceleration based on the fitting function of the thermalization curves. At the boundary separation $l=3$, the numeric curves for $w=0, -0.2, -0.4, -0.6, -0.8$ can be fitted as follows
\begin{eqnarray}
\begin{cases}
  -0.164186-0.00651677 t_0+0.0555242 t_0^2-0.0394575 t_0^3+0.184097 t_0^4-0.0993669 t_0^5\\
 -0.197916-0.00880886 t_0+0.119085 t_0^2-0.0138913 t_0^3+0.0682885 t_0^4-0.0466105 t_0^5\\
 -0.25763+0.00109052 t_0+0.302199 t_0^2-0.25667 t_0^3+0.167809 t_0^4-0.0538153 t_0^5\\
 -0.386678+0.16368 t_0+0.26886 t_0^2-0.319854 t_0^3+0.18971 t_0^4-0.0477778 t_0^5\\
 -0.775697+1.05582 t_0-1.13954 t_0^2+0.965802 t_0^3-0.414155 t_0^4+0.0665978 t_0^5
\end{cases}
\end{eqnarray}

\begin{figure}
\centering
\subfigure[$\texttt{}$]{
\includegraphics[scale=0.75]{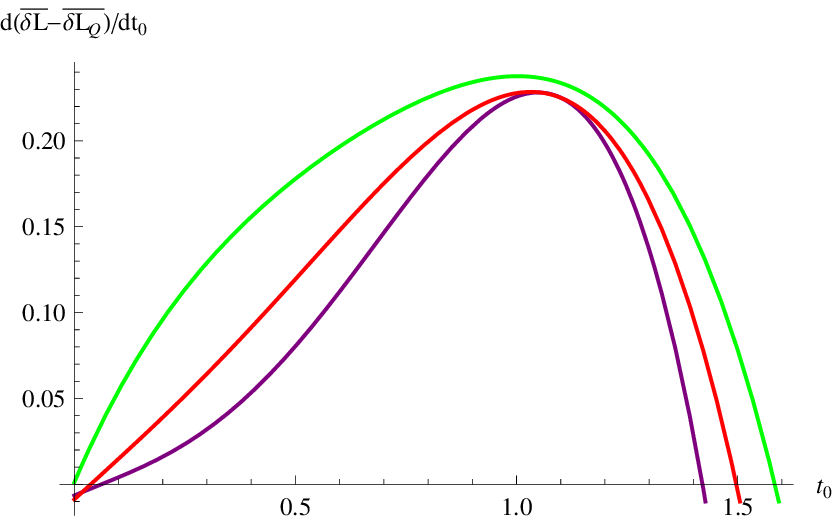} }
\subfigure[$\texttt{}$]{
\includegraphics[scale=0.75]{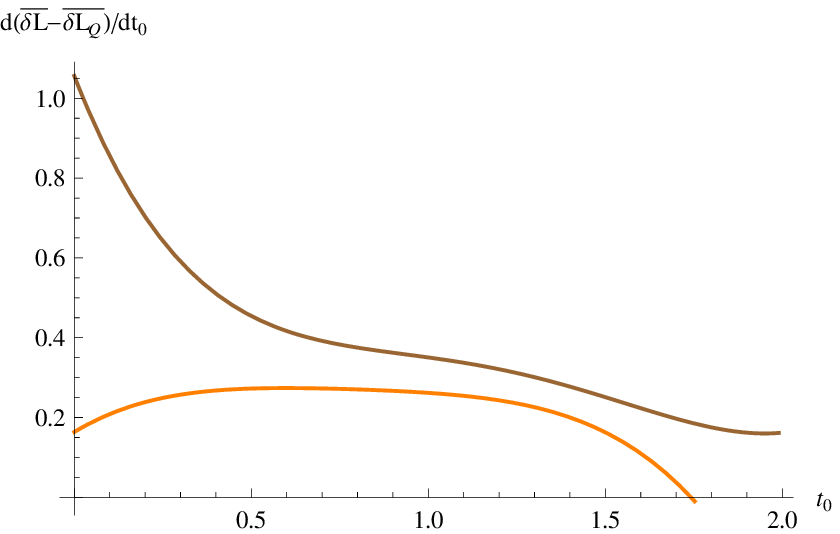} }
 \caption{\small Thermalization velocity of the renormalized geodesic  in a quintessence Vaidya AdS  black brane. The purple line, red line,  and green line in  (a) correspond to  $w=0, -0.2, -0.4$  and the orange line, brown line  in  (b) correspond to  $w=-0.6, -0.8$  respectively.} \label{fig5}
\end{figure}

\begin{figure}
\centering
\subfigure[$\texttt{}$]{
\includegraphics[scale=0.75]{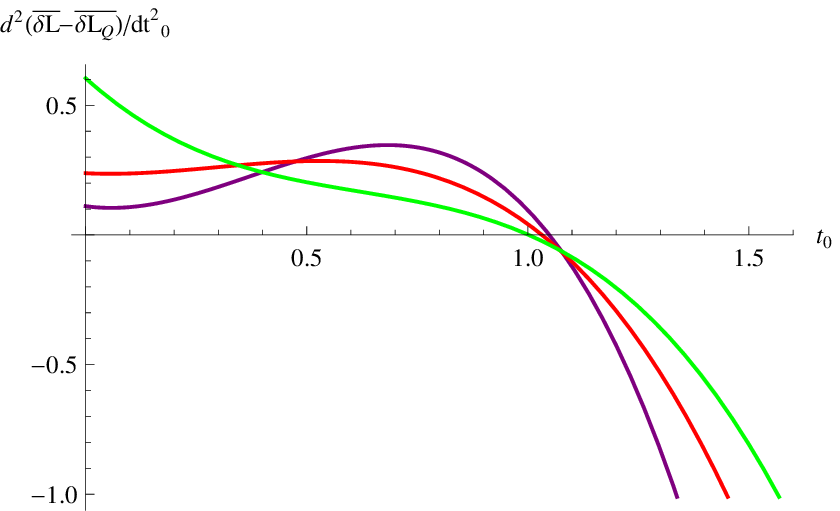} }
\subfigure[$\texttt{}$]{
\includegraphics[scale=0.75]{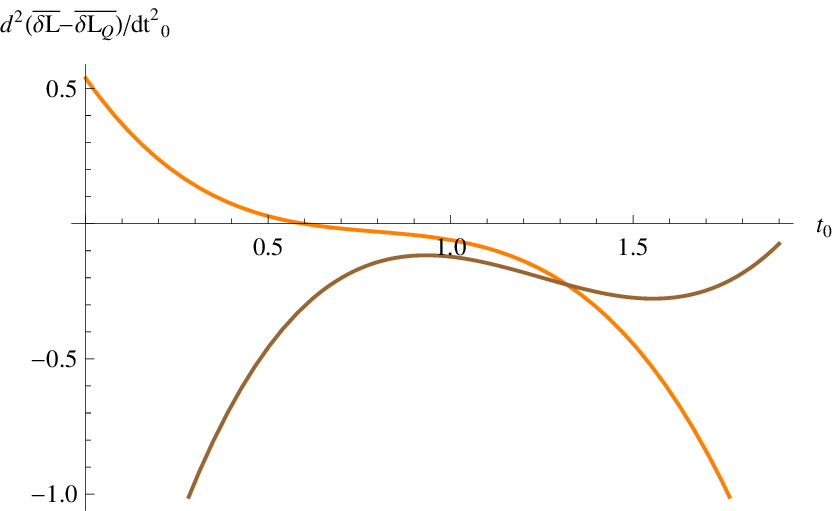} }
 \caption{\small Thermalization acceleration of the renormalized geodesic  in a quintessence Vaidya AdS  black brane. The purple line, red line,  and green line in  (a) correspond to  $w=0, -0.2, -0.4$  and the orange line, brown line  in  (b) correspond to  $w=-0.6, -0.8$  respectively.} \label{fig6}
\end{figure}

For small thermalization time, the functions are determined by the lower power of $t_0$, while for large thermalization time they are determined by the higher  power of $t_0$. Figure (\ref{fig4}) is the comparison of the numerical curves and fitting function curves. It is obvious that at
the order of $t_0^5$, the thermalization curve can be described well by the fitting function~\footnote{For higher order power of $t_0$, we find it has few contributions to the thermalization, including the thermalization velocity, thermalization acceleration and  the phase transition point which will be discussed next.}. With these functions, we can get the thermalization velocity, defined by $d(\overline{\delta L}-\overline{\delta L_{Q}})/dt$, and thermalization acceleration, defined by $ d^2(\overline{\delta L}-\overline{\delta L_{Q}})/dt^2$, which are plotted in Figure (\ref{fig5}) and Figure (\ref{fig6}).
From the thermalization velocity curves in Figure (\ref{fig5}), we see that there is a phase transition point at the middle stage of the thermalization for $w=0, -0.2, -0.4, -0.6$ respectively, which divides the thermalization into an acceleration phase and a deceleration phase. The phase transition point for different state  parameters can be read off from the null point of the acceleration curves in Figure (\ref{fig6}). It is easy to find that in the time range,
$0<t_0<1.042, 1.036, 0.9922, 0.6201$ for $w=0, -0.2, -0.4, -0.6$ respectively, the thermalization is in an acceleration process while for the other time range, it is in a deceleration process before it approaches to the equilibrium state. Furthermore, as the state parameter of quintessence decreases, the value of the phase transition point decreases. That is, smaller the state parameter is, earlier the thermalization decelerates. This result also indicates that the smaller state  parameter of quintessence delays the thermalization, which is consistent with the previous result. In addition, from (b) in Figure (\ref{fig5}), we find the thermalization is always decelerated and the acceleration in Figure (\ref{fig6}) is negative, which is different from other values of $w$, which accelerates first and then decelerates.

In addition, from the thermalization velocity curves in Figure (\ref{fig5}), we also find a non-monotonic behavior of the renormalized geodesic length for $w=0, -0.2$, which is indicated by the negative value of the renormalized geodesic length at the initial thermalization time. This non-monotonic behavior has also been observed and explained in~\cite{CK, Zeng2014, Zeng2015}. But in our paper, we find this non-monotonic behavior vanishes as the state parameter decreases. Especially for the case $w=-0.8$, the velocity of renormalized geodesic length in the vacuum state is so large that it should be decelerated in order to approach to the equilibrium state lastly.

\subsection{Wilson loop}

\begin{table}
\begin{center}\begin{tabular}{l|c|c|c|c|c}
 %\MC{3}{c}{\text{caption}}\\[5pt]
 \hline
% & \multicolumn{3}{c||}{MGCDM}   & \multicolumn{3}{c}{$\Lambda$CDM}  \\ \hline
%                             &        MGCDM        &                  &             &      $\Lambda$CDM    &                   & \\ \hline
% \MC{3}{|c|c|}{\ZZ{-8pt}{15pt}\hfill\normalsize   \hfill  \hfill\normalsize MGCDM     \hfill\normalsize $\Lambda$CDM  }\\ \hline
% \ZZ{-6pt}{22pt}
               &$w=0$ &          $w=-0.2$  &       $w=-0.4$   &       $w=-0.6$  &   $w=-0.8$    \\ \hline
$v_{\star}$=-0.777    & 0.554439    &  0.563105      &0.582772         & 0.633031        & 0.801127  \\ \hline
$v_{\star}$=-0.444    &0.898102    &0.925733         &0.975617        &1.0783     & 1.36818    \\ \hline
$v_{\star}$=-0.111     &1.17786      &1.22157      & 1.28985            &1.41595      &1.7401      \\ \hline
\end{tabular}
\end{center}
\caption{The thermalization time $t_0$ of the minimal area surface   probe  for different  state parameters of quintessence $w$ and different initial time $v_{\star}$ with the same boundary separation $l=1.6$.}\label{tab2}
\end{table}

%======================figure1====================

\begin{figure}
\centering
\subfigure[$w=-0.2$]{
\includegraphics[scale=0.75]{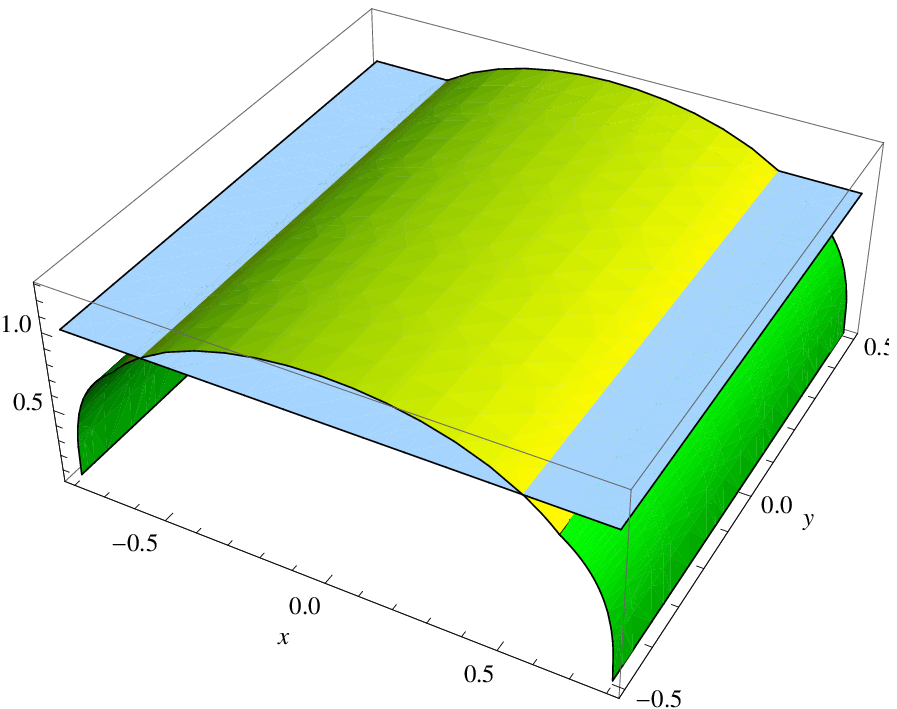}
 }
\subfigure[$w=-0.4$]{
\includegraphics[scale=0.75]{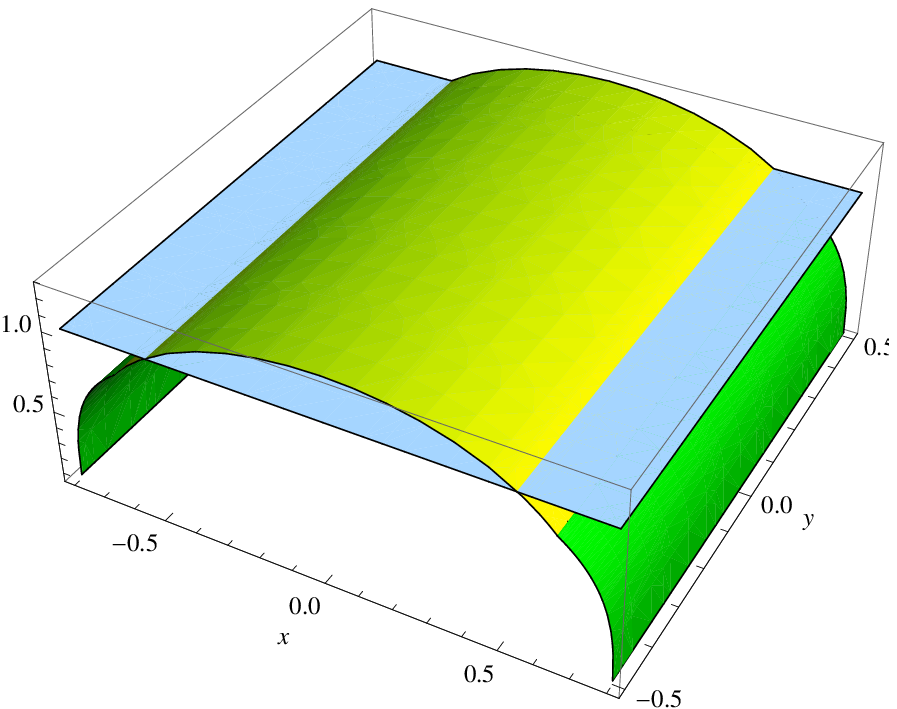}

}
\subfigure[$w=-0.6$]{
\includegraphics[scale=0.75]{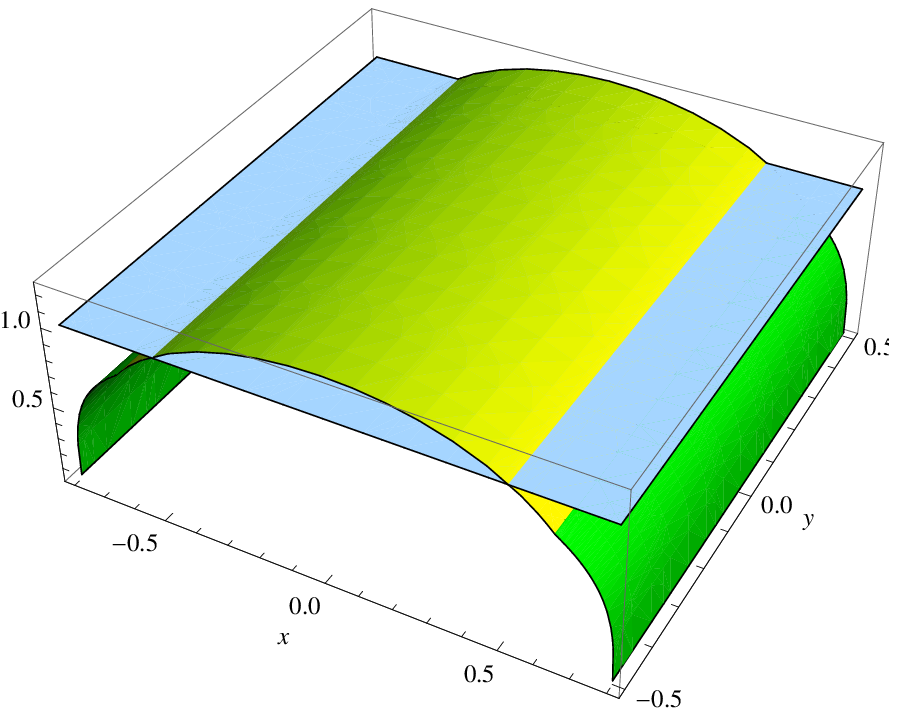}
}
\subfigure[$w=-0.8$]{
\includegraphics[scale=0.75]{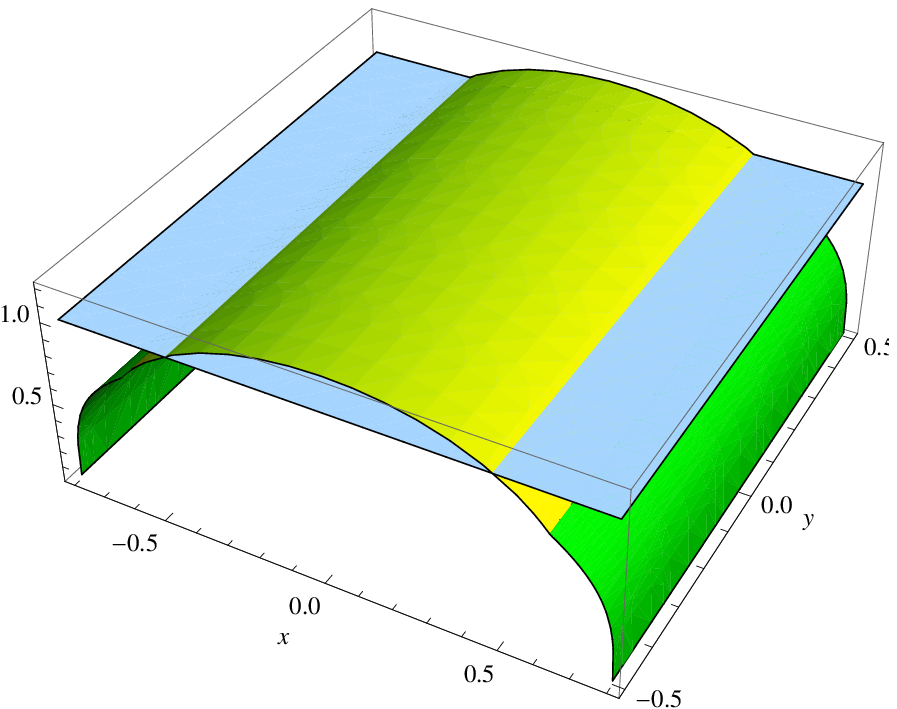}
}
 \caption{\small Motion profile of the minimal area surface   in the quintessence Vaidya AdS black brane. The boundary separation along the $x$ direction is $1.6$, and along the $y$ direction is $1$. The white surface is the location of the horizon.  The  position of the shell is  described by the junction between the  yellow surface and the green surface.} \label{fig7}
\end{figure}

~
\begin{figure}
\centering
\subfigure[$l=1$]{
\includegraphics[scale=0.75]{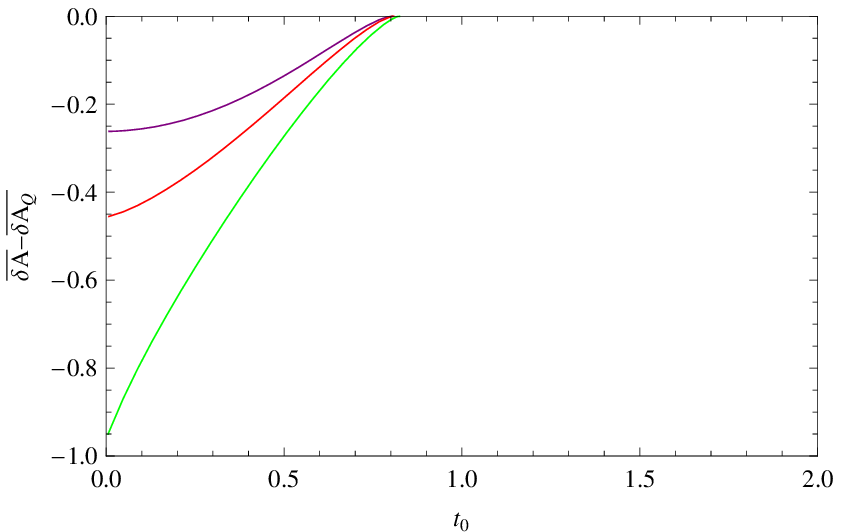} }
\subfigure[$l=1$]{
\includegraphics[scale=0.75]{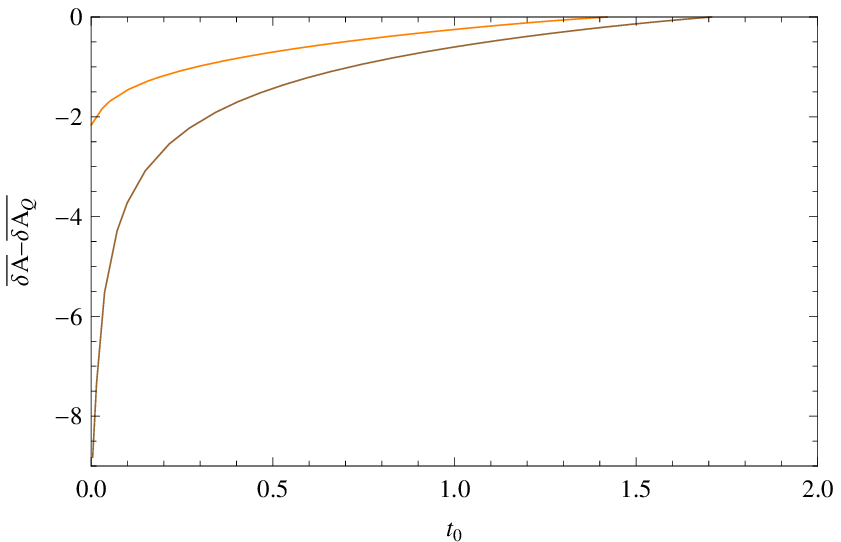}
}
 \caption{\small  Thermalization of the renormalized   minimal   area surface in a quintessence Vaidya AdS  black brane. The purple line, red line,  and green line in  (a) correspond to  $w=0, -0.2, -0.4$  and the orange line, brown line  in  (b) correspond to  $w=-0.6, -0.8$  respectively.} \label{fig8}
\end{figure}

\begin{figure}
\centering
\subfigure[$l=1.6$]{
\includegraphics[scale=0.75]{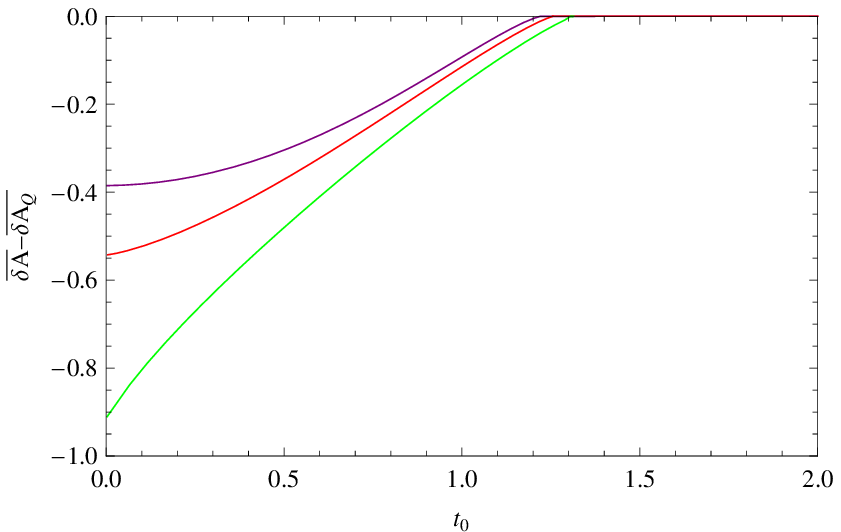} }
\subfigure[$l=1.6$]{
\includegraphics[scale=0.75]{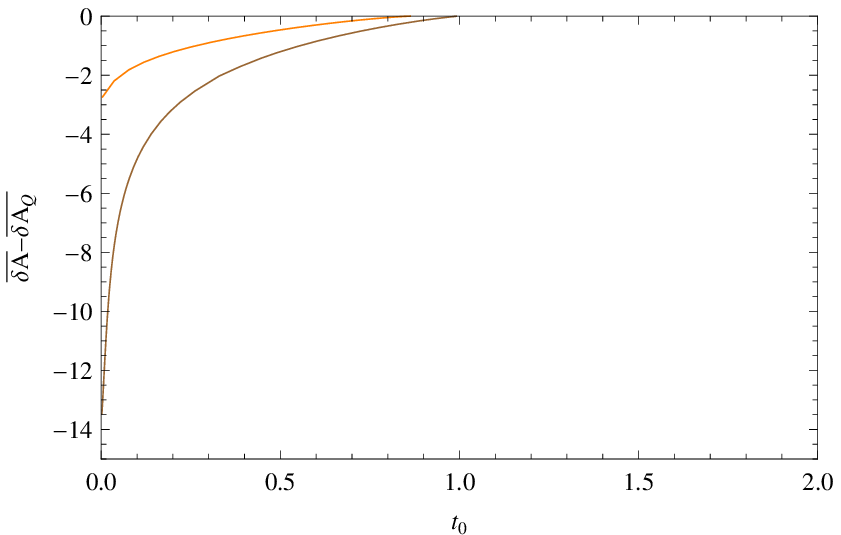}
}
 \caption{\small  Thermalization  of the renormalized  minimal   area surface  in a quintessence Vaidya AdS  black brane. The purple line, red line,  and green line in  (a) correspond to  $w=0, -0.2, -0.4$  and the orange line, brown line  in  (b) correspond to  $w=-0.6, -0.8$  respectively.} \label{fig9}
\end{figure}

With the same strategy, in this subsection we will use the Wilson loop to probe the thermalization of the quark gluon plasma. We should first solve the equations of motion of the minimal area surface, namely $n=2$ in Eq.(\ref{lequation}) and Eq.(\ref{aequation}), for different initial time and state parameters of quintessence. In Table (\ref{tab2}), we list the thermalization time for different state parameters at different initial time $v_{\star}$ with the same boundary separation. From it, we know that at a fixed initial time, as the state parameter decreases, the thermalization time decreases step by step, which implies that the smaller state parameters delay the thermalization. At $v_{\star}=-0.444$, we also plot the motion profiles of the minimal area surface for different state parameters, which are shown in Figure (\ref{fig7}). From (a) to (d) in Figure (\ref{fig7}), we know that as the state parameters decrease, the shell goes away from the horizon little by little. That is, for the smaller state parameters, the quark gluon plasma in the dual conformal field theory is harder to thermalize.
This behavior is the same as that observed in subsection~\ref{two_point}.

\begin{figure}
\centering
\subfigure[$l=3$]{
\includegraphics[scale=0.75]{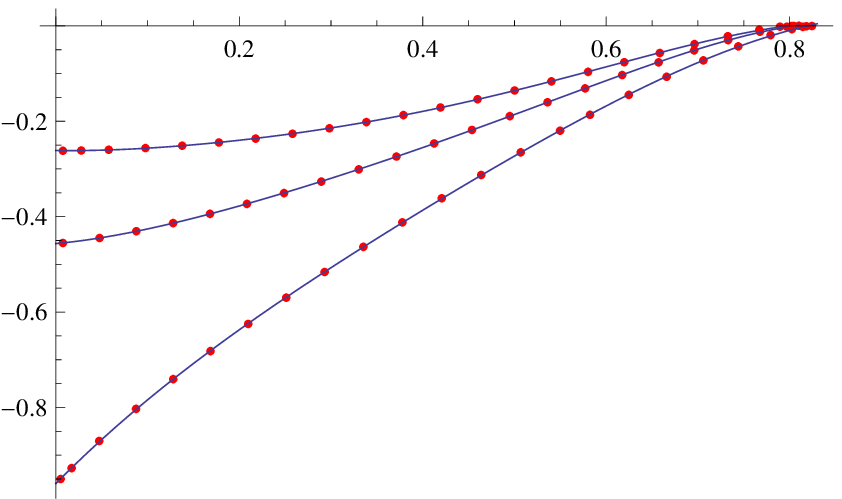} }
\subfigure[$l=3$]{
\includegraphics[scale=0.75]{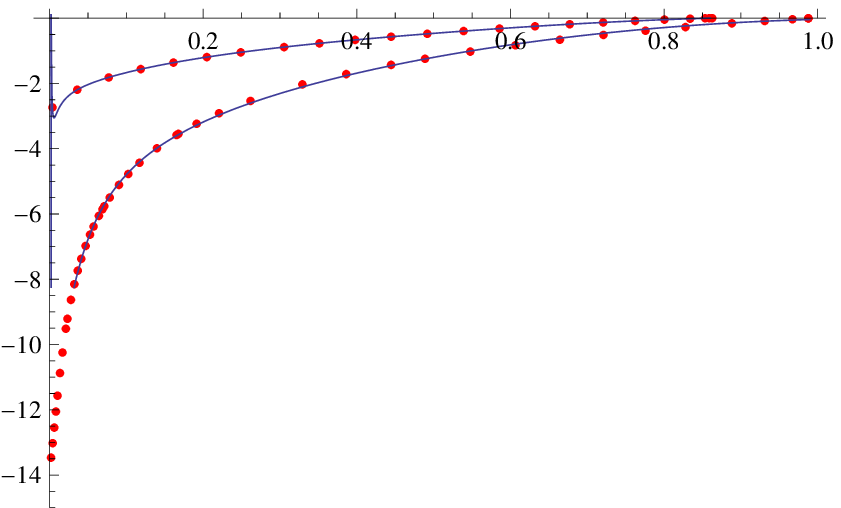}
}
 \caption{\small Comparison of the functions in Eq.(4.2) with the numerical results in  Figure (8).} \label{fig10}
\end{figure}

We further study the renormalized minimal area surface based on Eq.(\ref{aren}). Here we are interested in $l$ independent quantity $\overline{\delta A}-\overline{\delta A_{Q}}$, where $\overline{\delta A_{Q}}$ is the renormalized minimal area surface at the equilibrium state. Figure (\ref{fig8}) and Figure (\ref{fig9}) show the relation between the renormalized minimal area surface and thermalization time for different state parameters $w$ at different boundary separation. In each picture, the vertical axis indicates the renormalized minimal area surface while the horizontal axis indicates the thermalization time $t_0$.
From Figure (\ref{fig8}) and Figure (\ref{fig9}), we can observe the same phenomena when we consider the geodesic as a probe in the last subsection.
\begin{itemize}
\item At a fixed boundary separation, as the state parameter decreases, the thermalization time increases, which means that the smaller state parameters of quintessence delay the thermalization;
\item For the larger boundary separation, the transition of the renormalized minimal area surface between two adjacent state parameters is more obvious;
\item As the state parameters decrease, the renormalized minimal area surface at the initial thermalization time decreases. For the large state parameters, there is a slow evolution stage for the renormalized minimal area surface at the initial thermalization time, see (a) in Figure (\ref{fig8}) or Figure (\ref{fig9}). While for the small state parameters the renormalized minimal area surface enhances drastically at the initial thermalization time, see (b) in Figure (\ref{fig8}) or Figure (\ref{fig9}).
\end{itemize}

To understand the evolution of the renormalized minimal area surface profoundly, we will also get the fitting functions for different $w$. At the boundary separation $l=1$, the numerical curves for $w=0, -0.2, -0.4, -0.6, -0.8$  can be fitted as

\begin{figure}
\centering
\subfigure[$\texttt{}$]{
\includegraphics[scale=0.75]{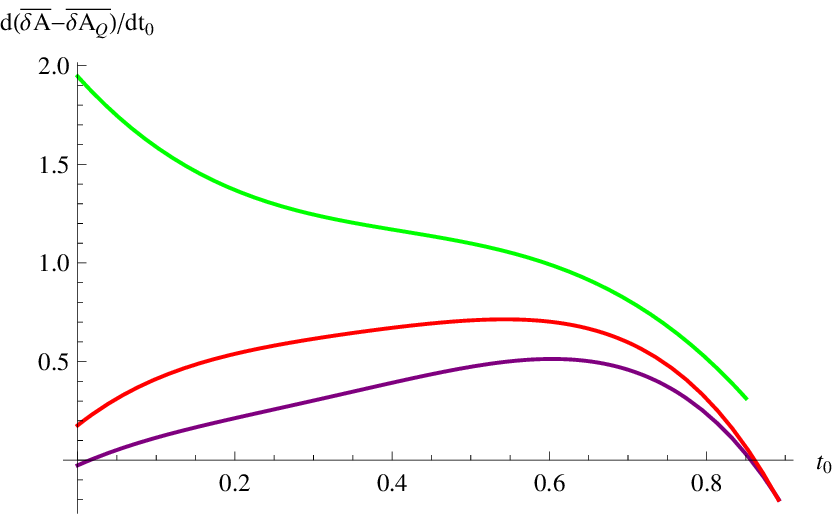} }
\subfigure[$\texttt{}$]{
\includegraphics[scale=0.75]{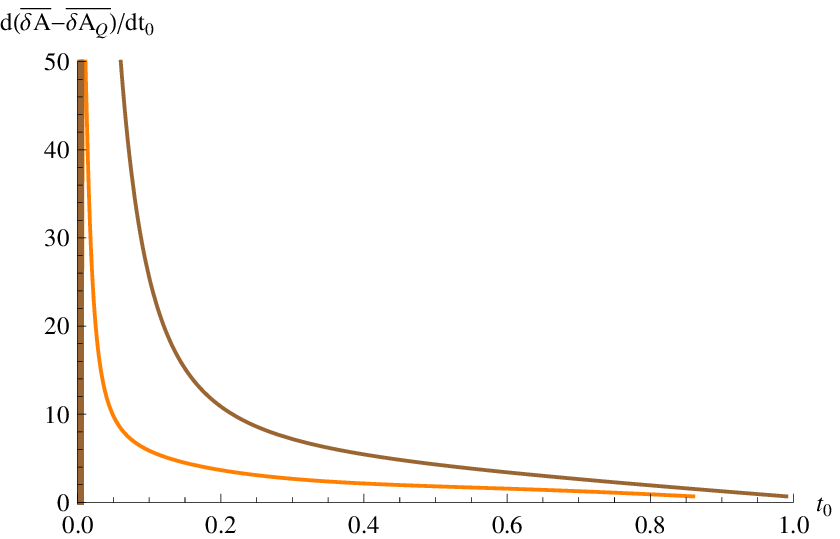} }
 \caption{\small Thermalization velocity of the renormalized   minimal area surface  in a quintessence Vaidya AdS  black brane. The purple line, red line,  and green line in  (a) correspond to  $w=0, -0.2, -0.4$  and the orange line, brown line  in  (b) correspond to  $w=-0.6, -0.8$  respectively.} \label{fig11}
\end{figure}

\begin{figure}
\centering
\subfigure[$\texttt{}$]{
\includegraphics[scale=0.75]{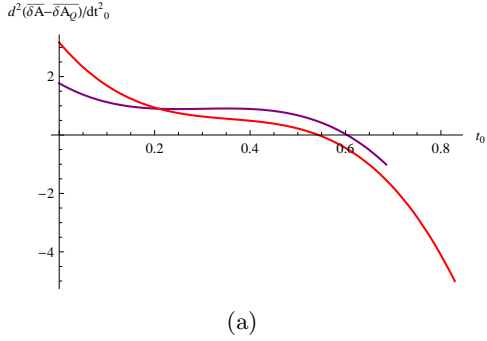} }
 \caption{\small Thermalization acceleration of the renormalized  minimal area surface  in a quintessence Vaidya AdS  black brane. The purple line and red line correspond to  $w=0, -0.2$.} \label{fig12}
\end{figure}

\begin{eqnarray}
\begin{cases}
  -0.261317-0.0260113 t_0+0.886079 t_0^2-1.55218 t_0^3+2.66284 t_0^4-1.77218 t_0^5 \\
 -0.456451+0.176607 t_0+1.58415 t_0^2-3.22188 t_0^3+4.19658 t_0^4-2.34786 t_0^5 \\
 -0.95938+1.94602 t_0-2.22108 t_0^2+3.13695 t_0^3-2.04563 t_0^4+0.141562 t_0^5 \\
 -2.14421+\frac{0.000033}{t_0^2}-\frac{0.011285}{t_0}+7.08951 t_0-14.1247 t_0^2+19.804 t_0^3-14.6582 t_0^4+4.10134 t_0^5 \\
 -3.02883+\frac{4*10^{-13}}{t_0^6}-\frac{5*10^{-10}}{t_0^5}+\frac{2*10^{-7}}{t_0^4}-\frac{0.000041}{t_0^3}+
 \frac{0.00436829}{t_0^2}-\frac{0.275087}{t_0}+6.18016 t_0-2.91635 t_0^2
\end{cases}
\end{eqnarray}

Figure (\ref{fig10}) is the comparison of the numerical results and fitting functions. We can see that the thermalization curves can be described well by the fitting functions besides the initial thermalization time for the case $w=-0.8$\footnote{For this part, we find it is hard to find a proper function to fit it. }. With these functions, we also can get the thermalization velocity and thermalization acceleration for different $w$, which are plotted in Figure (\ref{fig11}) and Figure (\ref{fig12}) respectively.

From the thermalization velocity curves in Figure (\ref{fig11}), we find for the cases $w=0, -0.2$, there is a phase transition point which divides the thermalization into an acceleration phase and a deceleration phase. The phase transition points can also be read off from the null point of the
acceleration curves, which are plotted in Figure (\ref{fig12}). It is obvious that in the time range, $0<t_0<0.6001$ for $w=0$ and $0<t_0<0.5417$ for $w=-0.2$, the thermalization is accelerated while for the other time range, it is decelerated before it approaches to the equilibrium state. Obviously, as the state parameter of quintessence decreases, the value of the phase transition point is smaller. For $w=-0.4, -0.6, -0.8$, we find the whole thermalization is decelerated. Especially for $w=-0.6, -0.8$, the thermalization decelerates drastically at the initial thermalization time and then decelerates smoothly. For all the state parameter of quintessence, we find the initial velocity of renormalized minimal area surface increases as the state parameter decreases. So for the case $w=-0.4, -0.6, -0.8$, the thermalization velocity of minimal area surface in the vacuum state is so large that it should be decelerated in order to approach to the equilibrium state lastly. Obviously the result obtained from the fitting functions of renormalized minimal area surface is similar to that obtained from the evolution of the renormalized geodesic length.

\section{Conclusions and Discussions}
Effects of the quintessence dark energy on the holographic thermalization have been investigated in this paper. The model in the bulk is realized by the collapsing of a shell of dust that interpolates between a pure AdS and a quintessence AdS black brane. The spacetime structure of the quintessence AdS black brane is found to be dependent on the state parameters of the quintessence dark energy $w$ with the variation range $-1\leq w<0$. As $w$ approaches to $0$, the spacetime reduces to the usual planar Schwarzschild AdS black hole and as $w$ approaches to $-1$, it reduces to a pure AdS space with a renormalized cosmological constant $1/L^2-M$. So for the large state parameters, namely near $w=0$, the thermalization behaves similar to that in the Vaidya AdS black branes with a slight delay in the onset of thermalization.  While for the small state parameters, namely near $w=-1$, the thermalization for both the thermalization probes has different behavior completely. That is, the delay in the onset of thermalization vanishes and to some case, it even increases drastically. The thermalization velocity in this case decreases always during the whole thermalization process, which is different from the one for large state parameters where the thermalization is divided into an acceleration phase and deceleration phase. The thermalization behaves differently for different state parameters of quintessence in that the state parameters affect the spacetime structure, which further reflects in the Hawking temperature in Eq.(\ref{temperature}). As stressed in \cite{Balasubramanian2}, the thermalization only becomes fully apparent at distances of the order of the thermal screening length $\tilde{l}_D\sim(\pi T)^{-1}$, where $T$ is the temperature of the dual conformal field. From (\ref{temperature}), we know that as $w\rightarrow 0$, the temperature is finite thus there is a delay in the onset of thermalization, while as  $w\rightarrow -1$, the temperature is infinitesimal, the delay vanishes and the thermalization begins immediately.

Another interesting phenomenon is that our model also can describe the transition for an vacuum state to another vacuum state. As $w$ approaches to $-1$, we know that the spacetime reduces to a pure AdS space with a renormalized cosmological constant $1/L^2-M$, which in the dual conformal field theory is dual to an vacuum state. Thus for the case $w\rightarrow -1$, the thermalization behavior also can describe the transition of the vacuum state. We will discuss this topic more specifically in the future.

%======================figure1===================

\section*{Acknowledgements}
We are grateful to Hongbao Zhang for   his  various valuable suggestions about this work.
 This work is supported  by the National
 Natural Science Foundation of China (Grant No. 11405016 and No. 11205226).

\end{document}